\documentclass{aa}
\usepackage{graphicx}
\usepackage{aalongtable}
\begin{document}
\title{A comparison among LBGs, DRGs and BzK galaxies: their contribution to
the stellar mass density in the GOODS-MUSIC sample}

   \author{
	  A. Grazian \inst{1}
	  \and
          S. Salimbeni \inst{1}
	  \and
          L. Pentericci \inst{1}
	  \and
	  A. Fontana \inst{1}
	  \and
          M. Nonino \inst{2}
	  \and
          E. Vanzella \inst{2}
	  \and
          S. Cristiani \inst{2}
          \and
          C. De Santis \inst{1}
          \and
          S. Gallozzi \inst{1}
	  \and
          E. Giallongo \inst{1}
          \and
          P. Santini \inst{1}
}

   \offprints{A. Grazian, \email{grazian@mporzio.astro.it}}

\institute{INAF - Osservatorio Astronomico di Roma, Via Frascati 33,
I--00040, Monteporzio, Italy
\and INAF - Osservatorio Astronomico di Trieste, Via G.B. Tiepolo 11,
I--34131, Trieste, Italy}

   \date{Received 10 July 2006 ; accepted 22 December 2006}
   \titlerunning{LBGs, DRGs and BzK in GOODS-MUSIC}

  \abstract
{
The classification scheme for high redshift galaxies is complex
at the present time, with simple colour selection criteria
(i.e. EROs, IEROs, LBGs, DRGs, BzKs), resulting in ill-defined
properties regarding stellar mass and star formation rate for these
distant galaxies.
}
{
The goal of this work is to investigate the properties of
different classes of high-z galaxies, focusing in particular on 
the Stellar Masses of LBGs,
DRGs and BzKs, in order to derive their contribution to the total mass
budget of the distant Universe.
}
{
We have used the GOODS-MUSIC catalog, containing $\sim$3000
Ks--selected ($\sim$10000 z--selected) galaxies with multi-wavelength
coverage extending from the U band to the Spitzer $8 \mu$m band,
with spectroscopic or accurate photometric redshifts. We have selected samples
of BM/BX/LBGs, DRGs and BzK galaxies, discussed the overlap and the limitations
of these criteria, which can be overcome with a selection criterion based on
physical parameters. We have then measured the 
stellar masses of these galaxies and
computed the Stellar Mass Density (SMD) for the different samples up to
redshift $\simeq 4$.
}
{
We show that the BzK-PE criterion is not optimal to select early type 
galaxies at the  faint end. BzK-SF, on the other hand, is highly contaminated
by passively evolving galaxies at red $z-Ks$ colours. We find that LBGs
and DRGs contribute almost equally to the global SMD at $z\ge 2$ and in
general that star forming galaxies form a substantial fraction of the
universal SMD. Passively evolving galaxies show a strong negative density
evolution from redshift 2 to 3, indicating that we are witnessing the
epoch of mass assembly of such objects.
Finally we find indications that pushing the selection to deeper magnitudes,
the contribution of  less massive DRGs could
overcome that of LBGs. Deeper surveys, like the HUDF, are required
to confirm this suggestion.
}
{
}


   \maketitle


\section{Introduction}

Until recently, the Lyman drop-out technique was the only selection 
criterion used to identify large 
numbers of galaxies at $z\ge 3$ in deep optical surveys
(\cite{madau,steidel}). This technique is effective in finding star
forming galaxies, characterised by a typical blue spectrum with
reduced dust absorption,
the so called Lyman Break Galaxies (LBGs).  In recent years, with the
availability of large IR detectors at 4-8 meter class telescopes, the
ISO and Spitzer data in the medium IR and the SCUBA observations, new types
of high-z galaxies have been identified, with spectral energy
distributions (SED) and physical
properties (Age, Mass, SFR, dust content) different from the
``classical'' LBG population.  Among them, it is
worth reminding the sub-mm galaxies (\cite{smail}), the distant red
galaxies (DRGs, \cite{franx,vandokkum}), Extremely Red Objects (EROs,
\cite{mcarthy,daddieros}), IRAC Extremely Red Objects (IEROs,
\cite{yan}) and the ``BzK'' galaxies (\cite{bzk}). The aim of these
different selection criteria is to find objects that could be missed
by the Lyman Break technique, due to the presence of an evolved
stellar population (dead and massive objects) or a young stellar
population heavily obscured by dust at high redshift (dusty
starbursts).

While the LBG selection identifies galaxies according to their less
obscured star formation properties based on UV rest frame colours,
EROs, DRGs, IEROs and BzK selection criteria include also optical rest
frame colours, which are less affected by recent SF episodes, but are
much more sensitive to the light coming from lower-mass
longer-lived stars. Another advantage of the selection from the
near-IR is that the rest frame optical light is less affected by dust
extinction.  However it is not yet clear how these different types of
galaxies overlap and what are the galaxies that dominate the number
counts and the mass density budget at high redshift.

Using different SFR indicators, \cite{reddy} and
\cite{reddy06} discussed the properties of LBGs, DRGs and
BzK galaxies in the redshift range $1.4\le z\le 2.6$ from the
GOODS-North survey,
and found a strong overlap (70-80\%) between LBG and BzK
galaxies, when the analysis is restricted to a Ks--selected sample.

LBGs, DRGs and BzKs present a similar SFR distribution as a
function of the Ks magnitude, the M/L ratio of DRG and BzK
galaxies (both massive and dusty) are larger than LBGs, but
their stellar mass does not exceed
the range spanned by optically selected galaxies. Previous
studies (\cite{adelberger04,shapley}) point to a similarity in the
metallicities, clustering and stellar masses of $Ks\le 20(vega)$
optical and near-IR selected galaxies at $1.5\le z\le 2.5$.
 Indeed, \cite{reddy} concluded that the presence or absence of star
formation may be the only significant difference between optical and
near-IR selected massive galaxies ($M_{star}\ge 10^{11}M_{\odot}$),
and the difference in SFR may be temporal, i.e. due to a phenomenon of
transient SF activity.
On the other hand, \cite{quadri} have studied the clustering properties
of K-selected galaxies at $2\le z\le 3.5$ in the MUSYC survey and
found that their correlation
length does not depend on the K band magnitude, but it increases strongly both
with $J-K$ and with $R-K$ colours. This suggests that K-bright
blue galaxies and K-bright red galaxies are fundamentally different,
at least in the clustering properties and hence on their host halo mass.

Recently, \cite{vandokkum06} compared the DRG and LBG selection
criteria at $2\le z\le 3$ using a K-selected sample ($Ks\le 21.3(vega)$
corresponding to $M\ge 10^{11}M_{\odot}$), finding that DRGs make
up 69\% (77\%) of the sample by number (by mass) and LBGs only 20\%
(17\%), with a small overlap between the two groups.
The census of the stellar mass density at high redshifts and faint
luminosities however is not well sampled in the recent literature 
as the SFR is, probably due to the lack of a comprehensive
database for LBGs, DRGs and BzKs, where accurate redshift measurements
and stellar mass estimates are available.

The GOODS-MUSIC sample (\cite{grazian06a}) allows to investigate the
role of different galaxy types (LBGs, DRGs and BzKs) in the universal
stellar mass budget at high redshift. Using this sample we are able to
select on the same wide, deep area all galaxy types, verify the
effectiveness of different colour criteria in sampling galaxies at
high redshifts, study the overlap within the sub-samples and their
contribution to the total stellar mass density of the Universe. In
this field, the stellar mass function for all galaxy types are already
studied (\cite{drory,caputi,massgoods}), the latter taking advantage
of the IRAC-Spitzer data, which are fundamental to derive the stellar
mass at higher-z (\cite{papovich05,massgoods}).  We focus here only on
two simple but largely debated questions, i.e. the role of LBGs, DRGs and
BzKs to the total stellar mass density of the Universe at $z\ge 1.4$, 
and the comparison between optically selected and near-IR selected samples.

The paper is organised as follows. In Sect.2, we remind the basic
feature of our GOODS-MUSIC dataset. In section 3 we describe the selection
criteria used to identify LBGs, DRGs and BzK galaxies, the overlap between
the samples, the observational limitations and their intrinsic
properties. In Section 4, we present the basic results of our analysis,
namely the resulting mass density per galaxy type, and its 
redshift evolution. In
Section 5 we discuss the results and derive our conclusions.

All magnitudes are in the AB system (except where otherwise stated)
and we adopt the
$\Lambda$-CDM concordance cosmological model ($H_0=70$, $\Omega_M=0.3$ and
$\Omega_{\Lambda}=0.7$).


\section{The Data}

In this work we use the GOODS-MUSIC (GOODS MUlticolour Southern
Infrared Catalog) sample, a 14 bands multicolour catalog extracted
from the deep and wide survey conducted over the Chandra Deep Field
South (CDFS, \cite{giavalisco}), in the framework of the GOODS
public survey.
The data comprise a combination of images that extend from U to
8.0$\mu$m, namely U-band data from the 2.2ESO ($U_{35}$ and $U_{38}$) and
VLT-VIMOS ($U_{VIMOS}$), the $F435W$, $F606W$, $F775W$ and $F850LP$
(Z-band) ACS images, the $JHKs$ VLT data and the Spitzer data provided by
the IRAC instrument at 3.6, 4.5, 5.8 and 8.0 $\mu$m. The total area of the
GOODS-MUSIC database results from the overlap of ACS $F435W$ band and Ks ISAAC
images, for a total of 143.2 sq. arcmin., which is covered by 12 bands,
most notably the Spitzer ones ($U_{35}U_{38}BVizJKs$ and IRAC). Only the
U-VIMOS and H bands have limited extension, of 90.2 and 78.0 sq. arcmin.,
respectively.

Since the detection mosaics have a complex, inhomogeneous depth and
are quite different in the $z$ and in the $Ks$ bands, we have divided
the whole sample into 6 independent catalogs for the $z$ band and in 6
for the Ks band, each with a well defined magnitude limit and area,
that we use to compute the stellar mass densities and other
statistical properties in this paper.  The limiting magnitudes for
each area are reported in Table 2 of \cite{grazian06a} and refer to a
90\% completeness.  The typical magnitude limit for most of the sample
is about $Ks=23.5$ and $z=26.0$ (corresponding to an area of 72 and 99
sq. arcmin., respectively), and extends down to $Ks=23.8$ and
$z=26.18$ in limited areas, as described in \cite{grazian06a}.

The GOODS-MUSIC database contains both a $z$--selected and a
$Ks$--selected catalog.  Colours have been measured using a specific
software for the accurate ``PSF--matching'' of space and ground based
images of different resolution and depth, that we have named ConvPhot
(De Santis et al. 2006). We have cross correlated our catalog with the
whole spectroscopic catalogs available to date, from a list of
surveys, assigning a spectroscopic redshift to more than 1000 sources.
In this work we use a spectroscopic sample that is wider than that
presented in \cite{grazian06a}, thanks to the increased number of
spectra publicly available (\cite{vanzella06}).  Finally, we have
applied our photometric redshift code, developed and tested over the
years in a series of works (\cite{fontana00}, \cite{cimatti02},
\cite{fontana03}, \cite{fontana04}, \cite{giallongo05},
\cite{massgoods}), that adopts a standard $\chi^2$ minimisation over a
large set of templates obtained from synthetic spectral models. The
comparison with the spectroscopic sample (\cite{grazian06a},
\cite{grazian06b}) shows that the quality of the resulting photometric
redshifts is excellent, with a r.m.s. scatter in $\Delta z/(1+z)$ of
0.03 and no systematic offset over the whole redshift range $0<z<6$.
The procedures that we adopted to extract this catalog and derive the
photometric redshifts are described at length in \cite{grazian06a}.

The final samples that we adopt here consist of 2931 galaxies down to
$Ks\simeq 23.8$ and 9862 galaxies down to $z=26.18$, taking into
account the complex magnitude limits of the survey.  On the
total $Ks$--selected sample, 815 galaxies (973 for the $z$--selected)
have reliable spectroscopic redshifts and the remaining fraction have well
trained photometric redshifts.

AGNs and stars are identified first from spectroscopic information.
We then distinguish galaxies from stars and AGNs using
morphological and photometric information, when spectroscopic data
are not available. Point-like sources are selected using
the star/galaxy separation flag (s/g) provided by SExtractor (\cite{sex})
in the $z$ band. We tune the selection on known spectroscopic stars,
as described in \cite{grazian06a}.
We use photometric information to further check
this criterion, using in particular the ``BzK'' colour criteria of
\cite{bzk}.

For the purposes of the present work, we will use the $Ks$--selected
catalog to study the properties of DRGs and BzK galaxies, while the
LBG sample is derived from the $z$--selected catalog. When discussing
the properties of LBGs, we restrict the GOODS-MUSIC sample only to the
area covered by deep U-VIMOS observations, since LBGs are selected
by means of U, V and I bands. In practise, for LBGs, we
use the area of 90.2 sq. arcmin. with U-VIMOS coverage, while for DRGs
and BzK galaxies, we use the total GOODS-MUSIC area (143.2 sq. arcmin.).


\section{Similarities and differences between LBGs, DRGs and BzK galaxies}

We will now compare the various colour criteria
used to find  high-z galaxies. In particular we will discuss the
similarities and differences between LBGs, DRGs and BzK galaxies,
to shed light on two important questions, namely the
comparison between $z$--selected and $Ks$--selected samples of galaxies
at high redshift, and the relative fraction of different galaxy types
in a $Ks$--selected sample of galaxies.

We remark here that these simple colour criteria (LBGs, DRGs and BzKs)
cannot distinguish between passively evolving and evolved galaxies.  A
passively evolving galaxy is defined as an object where SF has ceased by a time
much longer than the typical duration of the SF episodes, while an
evolved galaxy has a large population of old stars but can also have
recent/on-going star formation.  Usually, through simple colour
criteria, it is possible only to check for the presence of evolved
stellar populations.  If the wavelength baseline is extended, as in
the case of the GOODS database, it is possible to isolate also true
passively evolving galaxies.

\subsection{Selection of LBGs, DRGs and BzK galaxies}

An efficient method to select unobscured and modestly obscured
star forming galaxies at high
redshifts is the Lyman break technique, that is
effective at $2.8\le z\le 3.7$, as defined originally by \cite{sh93,madau,
steidel95,steidel99}.
A recent extensions at lower redshift are the so called
``BX'' galaxies, at $2.2\le z\le 2.8$ and ``BM'' galaxies, at $1.4\le
z\le 2.2$ (\cite{adelberger04}). Since the filter set
of the GOODS-MUSIC sample is different from the classical
$UGR$ adopted by Steidel and collaborators (\cite{steidel}), we have
used the following criteria, tuned by a comparison to the photometric
redshift distributions:

\begin{eqnarray}
-0.2 \le V-I \le 0.35\nonumber\\
U-V \ge 0.75(V-I)+1.15
\end{eqnarray}
for LBGs,

\begin{eqnarray}
U-V \ge 0.65(V-I)+0.25\nonumber\\
U-V < 0.75(V-I)+1.15\nonumber\\
-0.2 \le V-I \le 0.25
\end{eqnarray}
for ``BX'' and

\begin{eqnarray}
U-V \ge 0.65(V-I)\nonumber\\
U-V < 0.65(V-I)+0.25\nonumber\\
-0.2 \le V-I \le 0.45
\end{eqnarray}
for ``BM''.

We have also verified that these criteria are almost equivalent to the
original BM/BX/LBG criteria by reproducing for each galaxy in our catalog
the synthetic $UGR$ magnitudes of Steidel and collaborators using the best
fit spectrum derived by our SED fitting technique. The colour selections
used here are also consistent with criteria based on photometric redshifts.

The differences in the filter sets, colour selections and the fact that
the BM/BX/LBGs are traditionally selected in the $R$ band and not in the
$z$ band, as in this work, result in a redshift
distribution of the BM/BX/LBG galaxies wider than in \cite{adelberger04} and
\cite{reddy}. There is a small sample of BX and LBGs at
$z_{phot}\sim 0.4$, due to galaxies at lower redshifts with $U-V$ and
$V-I$ colours typical of high-z galaxies. In the following, however,
we study the properties of BM/BX/LBGs selected with our colour criteria
only at $z_{phot}\ge 1.4$, thus avoiding the contamination from these lower
redshifts galaxies. In addition, BM/BX/LBGs were originally relatively bright
galaxies, since the historical criterion used was $R\le 25.5$, while in this
work
the selection is pushed till the nominal completeness of the GOODS-MUSIC
sample. Adopting for our BM/BX/LBG sample a fainter limit in
a redder band, however, should not introduce further contaminants,
since these galaxies are selected for their typical blue optical colours.

When applied to the $Ks$--selected sample, these colour cut, with the
additional constrain $z\ge 1.4$ for consistency with the classical 
BM/BX/LBG criteria, selected 166 galaxies
with substantial star formation at $1.4\le z\le 3.7$ and $Ks\le 23.8$,
with U-VIMOS magnitude determination:
hereafter we will refer to this sample as BM/BX/LBG-Ks.  The $z$ band,
however, is much more appropriate to select star forming galaxies
(with low dust extinction), though it is not the best choice for the
stellar mass estimate, in particular at high redshifts. Using the
same criteria as above, we have therefore selected 1345 galaxies in
the $z$ band ($z\le 26.18$), and we will refer to this
sample as BM/BX/LBG-Z.

Recently, \cite{bzk} proposed a criterion to identify galaxies at
$1.4\le z\le 2.5$, differentiating them between passively evolving
(BzK-PE), selected according to the criterion
$BzK \equiv (z-K)_{AB}-(B-z)_{AB}< -0.2$ and $(z-K)_{AB}> 2.5$,
and star-forming galaxies (BzK-SF), which can be
effectively isolated from lower redshift interlopers and stars
through the simple
criterion $BzK> -0.2$. The advantage of the BzK criterion
is that this colour combination is insensitive to extinction since the
reddening vector is parallel to the $BzK=-0.2$ dividing line, as shown in
\cite{bzk}. This
technique reveals particularly efficient for sampling the dusty side of
star forming galaxies at intermediate/high redshifts.  Using the BzK
selection criteria, we have extracted 89 BzK-PE and 747 BzK-SF
galaxies down to the conservative $Ks$ completeness
limits of the survey, described in Section 2 and in \cite{grazian06a}.

Finally we focus on the so-called Distant Red Galaxies (DRGs, \cite{franx}).
These galaxies are selected through the $J-K>2.3$(vega) criteria, originally
designed to be sensitive to galaxies with large 4000 \AA~and/or Balmer
breaks at $z\ge 2$. These features become strong for ages larger than 1 Gyr,
even if they are also sensitive to the metallicity of galaxies
(\cite{poggianti,kauffmann}). The presence of a Balmer and/or 4000 \AA~break
can thus be an indication of an evolved stellar population.
It turned out also that this single colour selection
is sensitive both to evolved galaxies and to dusty
starbursts at $z\ge 1-2$, as shown in \cite{nmfs04,papovich05} and
\cite{grazian06b}. DRGs thus appear to be a mixed/composite sample.
The DRGs in the GOODS-South sample are already well studied
(\cite{papovich05,grazian06b}): we select 179 galaxies of this type
down to $Ks=23.8$, which span a large redshift range $1\le z\le 4$.

\begin{figure}
\includegraphics[width=9cm]{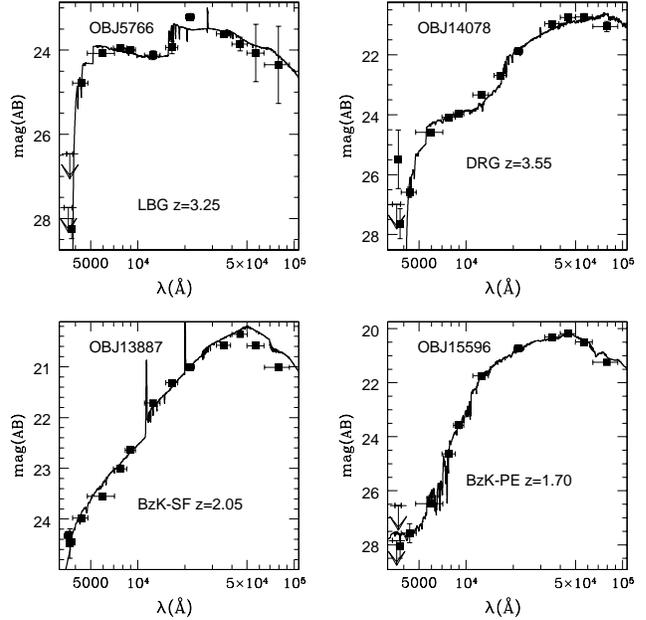}
\caption{
An example of LBG, DRG, BzK-SF and BzK-PE from the GOODS-MUSIC sample
(\cite{grazian06a}). The x-axis corresponds to the observed
wavelengths, while the y-axis refers to the observed magnitudes in
the AB photometric system.
}
\label{dimspec}
\end{figure}

In Fig. \ref{dimspec} we show typical SED  
of an LBG, a DRG, a BzK-SF and a
BzK-PE galaxy.

\subsubsection{The number density of LBGs, DRGs and BzK galaxies in the GOODS
South field}

The area of the GOODS South field is relatively large compared to the
deep pencil beam surveys like the two Hubble Deep Fields and the Hubble
Ultra Deep Field, but it is still limited and therefore subject to
cosmic variance, as demonstrated by the prominent peaks in the spectroscopic
redshift distribution (\cite{vanzella06}). It is also known that
the number density of X-ray sources in the GOODS-South is lower compared
to other deep X-ray surveys, i.e. the Chandra Deep Field North
at faint fluxes and hard energy band (2-8 keV, \cite{hasinger}),
and there are claims that this underdensity extends
to optically/NIR normal galaxies.

We have derived the number densities of BM/BX/LBGs, DRGs and BzK galaxies
in the GOODS South and compared them to the densities obtained by
large (but shallower) area surveys, in order to verify if GOODS-MUSIC
is a representative sample.

In their 0.81 sq. deg. survey, \cite{adelberger04} found
that at $R\le 25.5$ the number densities of BM, BX and LBGs are 3.28, 4.82
and 1.70 galaxies per sq. arcmin., respectively.
For comparison, in the GOODS-MUSIC sample our criteria for BM,
BX and LBGs select 3.35, 5.64 and 2.22 galaxies per sq. arcmin., at a limiting
magnitude of $Z\le 25.0$, $Z\le 25.3$ and $Z\le 25.3$\footnote{The $R-Z$ colour
of BM/BX/LBGs changes with redshifts, galaxies becoming redder at lower
redshifts.}, respectively. If we limit our BM/BX/LBGs at $z_{phot}\ge 0.5$
the number densities are 3.22, 4.40 and 1.70 galaxies per sq. arcmin. and
therefore agree much better with those of \cite{adelberger04}.

Recently, \cite{quadri} found 0.89 DRGs per sq. arcmin. at $Ks\le 21(vega)$;
in the GOODS-MUSIC sample at the same magnitude limit the number
density is similar, with 0.80 DRGs per sq. arcmin. Indeed, the DRG counts
in the GOODS South is consistent with that derived in deeper surveys, like
FIRES, as shown in \cite{grazian06b}.

The seminal work for the BzK selection criteria by \cite{bzk} is based on
the K20 survey (a subset of the GOODS-South field) and the number densities
of BzK-SF and BzK-PE at $K\le 20(vega)$ are 0.91 and 0.22 galaxies per
sq. arcmin., respectively. In a larger field, \cite{kong} found, at the same
magnitude limit, 1.20 $BzK-SF/arcmin^2$ and 0.38 $BzK-PE/arcmin^2$,
respectively.
In the GOODS-MUSIC sample, at the same magnitude limit, the BzK-SF galaxies
have a slightly lower number density (0.77), while the BzK-PE counts are
consistent (0.39) with the larger area survey of \cite{kong}.

An independent estimate of the number densities for BzK galaxies is provided
by \cite{reddy06}: in the GOODS-North the number densities are 3.1 and 0.24
$arcmin^{-2}$ for BzK-SF and BzK-PE, respectively, at $K\le 21(vega)$.
In the GOODS-MUSIC sample at this deeper magnitude cut, the densities are
comparable or slightly higher, with 3.2
and 0.65 galaxies per sq. arcmin. for BzK-SF and BzK-PE, respectively.
The large variation in number density for the BzK-PE sample is due probably
by the large angular clustering of this evolved population and the limited
area which is more sensitive to the field-to-field variations.

We thus confirm that the GOODS-South region is not underdense at
intermediate/high redshifts ($1\le z\le 4$) and, though affected by cosmic
variance, it is representative of the general properties of the
distant Universe.

\subsection{The overlap between samples}

\begin{figure}
\includegraphics[width=9cm]{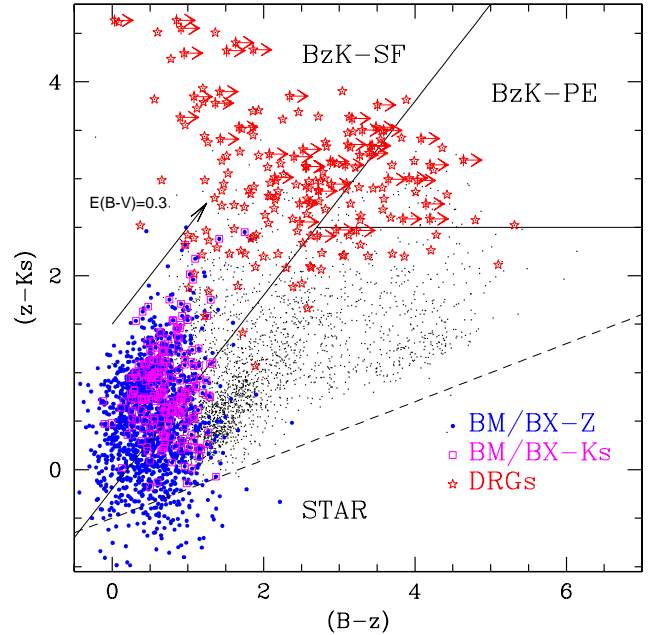}
\caption{
The $B-z$ vs $z-Ks$ colour for $Ks$--selected galaxies in the
GOODS-MUSIC sample (small dots). We do not plot AGNs or stars. Blue small
points show the star forming galaxies selected with the BM/BX
method (from the $z$--selected sample), magenta empty squares are the
Ks--selected BM-BX galaxies, while red stars are the DRGs.
Horizontal arrows indicate that the B magnitude estimation is
based on an upper limit at $1\sigma$.
The dashed line shows the separation between galaxies (up) and stars (down)
according to \cite{bzk}. The diagonal solid line at $BzK=-0.2$ allows to
isolate
star forming galaxies at $1.4\le z\le 2.5$ (up) from galaxies at lower
redshifts or from passively evolving one in the same redshift range
($z-Ks\ge 2.5$).  The reddening vector for $E(B-V)=0.3$ is plotted at the
left of the diagram, and is parallel to the $BzK=-0.2$
line. This ensures that the BzK-SF criterion is insensitive to dust
extinction.
}
\label{bzk}
\end{figure}

Fig. \ref{bzk} shows the distribution of BM/BXs and DRGs in the
$B-z$ vs $z-Ks$ diagram. In the BzK plot the loci of star-forming
(BzK-SF) and passively evolving galaxies (BzK-PE) are separated
by the $BzK=-0.2$ diagonal line.

\begin{table*}
\caption[]{Overlap between DRG, BM/BX and BzK galaxies at $1.4\le z\le 2.5$}
\begin{tabular}{lcccc|c}
\hline
\hline
Sample & DRG & BzK-PE & BzK-SF & BM/BX-Ks & BM/BX-Z \\
\hline
DRG & 78 & 5/78 (6\%) & 72/78 (92\%) & 2/78 (3\%) & 1/78 (1\%) \\
BzK-PE & 5/37 (14\%) & 37 & 0/37 (0\%) & 0/37 (0\%) & 0/37 (0\%) \\
BzK-SF & 72/327 (22\%) & 0/327 (0\%) & 327 & 123/327 (38\%) & 121/327 (37\%) \\
BM/BX-Ks & 2/124 (2\%) & 0/124 (0\%) & 123/124 (99\%) & 124 &
122/124 (98\%) \\
\hline
BM/BX-Z & 1/913 (0.1\%) & 0/913 (0\%) & 121/913 (13\%) & 122/913 (13\%) &
913 \\
\hline
\hline
\end{tabular}
\label{popgal}
\\
All the samples are restricted to the area covered by U-VIMOS in order
to select 
BM and BX galaxies and limited for all galaxy type to
$1.4\le z\le 2.5$ where the BzK criterion is efficient. The BM/BX-Ks sample
represents galaxies in the $Ks$--selected database which resemble BM or BX
colour criteria, while BM/BX-Z refers to the $z$--selected sample. The
diagonal of the matrix indicates the total number of galaxies selected for
each type. It is useful to remind that the BM/BX-Z sample is $z$--selected,
and thus cannot be compared directly with the $Ks$--selected samples, but
it is useful only as reference.
\end{table*}

In Table \ref{popgal} we report the overlap  between
different populations: these samples are restricted to the area covered by
U-VIMOS and limited for $Ks$--selected galaxy at $1.4\le z\le 2.5$, in order
to compare homogeneous classes. We add to
Tab.\ref{popgal} the BM/BX galaxies that are $z$--selected, for the sake of
completeness.

From Figure \ref{bzk} and Table \ref{popgal} we can draw the following
considerations.

Clearly, DRGs and BM/BXs are quite orthogonal as selection criteria, 
with a very small overlap of 
only 2 (out of 78 DRGs and 124 BM/BXs) galaxies in common between them.

On the other hand there is significant overlap between 
BzK and the BM/BX or DRG criteria:

\begin{itemize}
\item
The BzK-SF has overlap both to DRGs ( presumably their subset of dusty
star forming galaxies)
and BM/BXs ( less obscured star forming galaxies): we can
therefore conclude that 
it is a good selection criterion for star forming galaxies, and 
quite complete,  regardless of dust obscuration properties.

The fraction of BM/BX galaxies which are also BzK-SF is 99\%, while the
fraction of BzK-SFs that are also BM/BXs is 38\%.
Thus, BM/BX can be seen as a
subsample (less obscured) of typical starforming galaxies, represented
by the BzK-SF class.  This result is slightly comparable to that of
the GOODS-North field where \cite{reddy} found an overlap of $60-80\%$
between BM/BX galaxies and BzK-SF at $z\sim 2$, based on a
$K$--selected sample: the overlapping fraction also increases towards
fainter galaxies.

\item
The DRG population has overlap with all other selection criteria, with a
predominance of BzK-SF. Indeed, DRGs are a mix of 
two populations,  where star-forming galaxies strongly obscured by
dust and massive/evolved galaxies coexist, and possibly share both these
properties.
It is well known (\cite{papovich05,grazian06b}) that low-z DRGs are mainly
galaxies caught in their dusty starburst phase, which have thus distinct
SF properties in respect to passively evolving galaxies. The high-z DRG
subset, instead, is a mix of dusty starburst and evolved
massive galaxies, as shown by recent deep IR spectroscopy (\cite{kriek}).

Note however that a number of DRG galaxies that lay in the
upper left part of the BzK plot, only have limits in the B-z
colour. Therefore it remains unclear whether they would still lay in
the BzK-SF locus or migrate to the BzK-PE area if deeper B-band data
were added. In the next section we will discuss these observational
limitations in more detail.

\item
Among the BzK-PE galaxies, 14\% are also DRGs, as one would expect that
evolved galaxies are also selected by the Balmer and/or 4000 \AA~break
criterion.
However, a lot are not DRGs: from an inspection of their SED, we find that
these objects are all evolved galaxies, with a Balmer and/or 4000 \AA~break
at redshift smaller than 2, or with a $J-K$ colour just below the
defining DRG cut. Indeed, also \cite{reddy}, based on a spectroscopic
selected sample,
find  that the redshift distributions of BzK-PE galaxies and DRGs have very
little overlap.

The BzK-PE criterion is orthogonal to methods selecting star forming
galaxies, thus it is not contaminated by obscured starbursts mimicking the
colour of a red/evolved galaxy.
It is therefore an effective criterion to find
evolved galaxies at $z\sim 2$, although as we will discuss in
the next section, it is highly incomplete mainly due to observational
limitations.
\end{itemize}

\subsection{Observational limitations of LBG, DRG and BzK criteria}

In Figure \ref{bzk} there are a few star forming galaxies selected by the
BM/BX-Ks criteria which stand below the $BzK=-0.2$ line\footnote{
Clearly, the number of $z$--selected BM/BX galaxies below the colour cut
is much higher. This is
better shown in the colour figure where the BM/BX-Z and BM/BX-Ks samples
are clearly distinguished.}.
This seems to indicate
a small incompleteness of the BzK-SF selection towards star-forming
galaxies with low extinction, as confirmed by recent
VIMOS spectroscopy for a large ($>1000$) sample of galaxies
in the GOODS-South
field (\cite{popesso}). However, part of the effect is also due to the
different redshift interval of the two samples, $1.4\le z\le 2.5$ for the
BzK-SF and $1.4\le z\le 2.8$ for the BM/BX criterion.

As noted in the previous Section, most DRGs lay in the locus of BzK-SF
galaxies, with only a small fraction falling in the BzK-PE selection
area.  However when we analyse the galaxies individually we find
several DRGs with very red $z-Ks$ colour and upper limit in the $B$ band
in the locus of BzK-SF, whose SEDs are
better fit by an evolved galaxy at redshift $\ge 1.4$. In
agreement with Fig. 12 of \cite{reddy}, we find that in the BzK
diagram the locus proposed by \cite{bzk} for evolved
galaxies is not well defined, when considering galaxies with very red $z-Ks$
colours. A motivation for this effect is the lack of $B$ band deep
enough to provide a good diagnostic when very red galaxies in $z-Ks$
are considered. \cite{bzk} in fact used this photometric criterion on
the K20 sample, with Ks-band magnitude limited at $Ks=20(vega)$ and
taking advantage of the relatively deep B-band image provided by
GOODS. When fainter galaxies in the Ks-band are included, as in this
work, the BzK criterion starts to be ineffective in isolating early
type galaxies at high redshifts, as discussed also in \cite{reddy} and
\cite{renzini}.
In addition, few galaxies with detection in the B band and very red $z-Ks$
colour are also better fit by evolved galaxy SED, suggesting that
there is an intrinsic mix of star-forming and evolved galaxies at
$z-Ks\ge 2.5$.
This is predicted also by \cite{bzk}, who in Fig. 8 show that galaxies with
high formation redshift ($z_{form}\ge 5$) and relatively short e-folding time
of the star formation, at redshift $z\ge 2$ fall in the BzK-SF region
before migrating to the BzK-PE region at $z\sim 1.4$. We have checked some
of our evolved galaxies in the BzK-SF region and they have indeed
very high formation redshift and $z_{phot}\ge 2$ (or spectroscopic
when available).

\begin{figure}
\includegraphics[width=9cm]{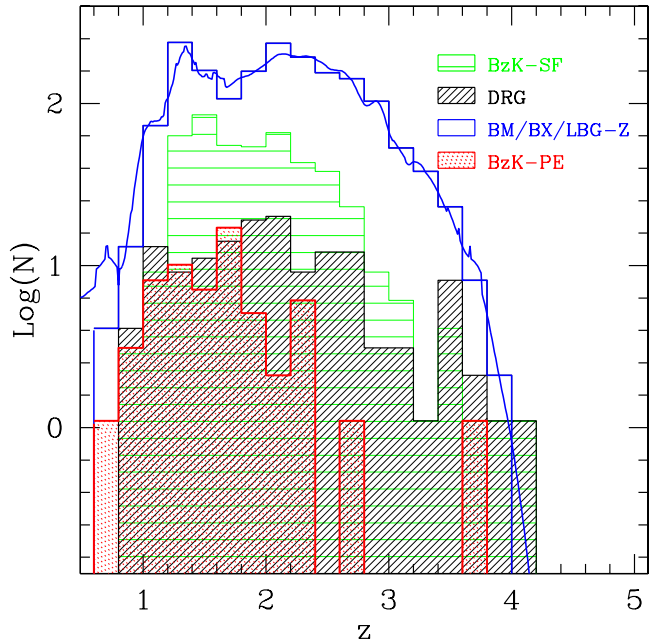}
\caption{Photometric redshift distribution of BM/BX/LBGs ($z$--selected),
DRGs, BzK-SF and BzK-PE galaxies, all limited to the area covered by U-VIMOS
observations.  We limit the BM/BX/LBG-Z sample at $z_{phot}\ge 0.6$ to
avoid the lower redshift contaminants selected by our criteria.
The continuous blue line is the sum of the probability distribution
function in redshift for each BM/BX/LBG galaxy, derived by our
photometric redshift code.
It is in agreement with the distribution using
the best estimate values for the photometric redshifts (solid histogram).
}
\label{zhist}
\end{figure}

In Fig. \ref{zhist} we show the photometric redshift distribution of BM/BX/LBGs
($z$--selected), DRGs, BzK-SF and BzK-PE galaxies, all limited to the
area covered by U-VIMOS observations. When available, spectroscopic redshifts
are used. In this plot there are BzK-SF galaxies at $z\ge 3$, even if the
BzK criterion should be efficient in the range $1.4\le z\le 2.5$. A
detailed analysis of the SED and physical parameters of these objects
indicates that they are almost all evolved galaxies with the Balmer
and/or 4000 \AA~break
falling between the J and the Ks band, thus resembling the DRG criterion.
In the BzK diagram they occupy the DRG locus with extreme colour in $z-Ks$.

The photometric redshift distribution shown in Fig. \ref{zhist} is quite
different from that shown by \cite{reddy}, since photometrically selected
DRGs and BzK-SFs show a wider redshift range than spectroscopically selected
ones.
These differences probably arise because of the limitation of the spectroscopic
redshifts, which are biased toward optically bright objects and
can depend on the instrumentation used to obtain the spectrum and to the
spectral type of the galaxy. In addition, part of the effect could be
due to field-to-field variance, which is not negligible even with the volumes
covered by the GOODS survey.

In summary, we have carried out a detailed comparison between the LBG,
DRG and BzK selection criteria. The selection of galaxies according to
the LBG criterion is sensitive only to the moderately obscured
SF galaxies, missing the dusty starburst objects (60\% of star forming
galaxies at $1.4\le z\le 2.5$). DRGs, instead, are less sensitive to dust
obscuration effects, but comprises a mix of two populations, the
old/evolved galaxies and the dusty starbursts at
intermediate/high redshifts. The BzK criterion is highly efficient in the
redshift range $1.4\le z\le 2.5$, but when galaxies start to become
faint, it is critical to distinguish between star-forming and evolved galaxies,
resulting in an underestimation of the passively evolving population.

\subsection{Intrinsic properties}

To overcome the limitations of LBG, DRG and BzK selection criteria, which are
based simply on observed colours, we use another approach based on
spectral fitting technique.

The spectral fitting technique adopted here is the same
that has been developed in previous papers
(\cite{fontana03,fontana04,grazian06a} and \cite{massgoods}), and similar
to those adopted
by other groups in the literature (e.g. \cite{dickinson03}, \cite{drory04}).
Briefly, it is based on the comparison between the observed multicolour
distribution of each object and a set of templates, computed with standard
spectral synthesis models (Bruzual \& Charlot 2003 in our case),
and chosen to broadly encompass the variety of star--formation histories,
metallicities and extinction of real galaxies.  To compare with
previous works, we have used the Salpeter IMF, ranging over a set of
metallicities (from $Z=0.02 Z_\odot$ to $Z=2.5 Z_\odot$) and dust
extinction ($0<E(B-V)<1.1$, with a Calzetti or a Small Magellanic Cloud
extinction curve). Details
are given in Table 1 of \cite{fontana04}. For each model of this grid, we have
computed the expected magnitudes in our filter set, and found the
best--fitting template with a standard $\chi^2$ normalisation. The
stellar mass and other best--fit parameters of the galaxy,
like SFR, age, $\tau$ (the star formation e-folding timescale),
metallicity and dust extinction, are fitted simultaneously
to the actual SED of the observed galaxy. The redshift of each galaxy is
fixed during the fitting process to the spectroscopic or photometric
redshift derived in \cite{grazian06a}.
Clearly, this approach requires the availability of spectroscopic or
photometric redshifts of good quality, in order to derive precisely
rest frame properties for these galaxies.

We define passively evolving galaxies according to the physical criterion
$age/\tau\ge 4$.
This quantity is in practice the inverse of the Scalo parameter
and a ratio of 4 is chosen as an arbitrary value to distinguish galaxies with
pure evolved stellar populations from galaxies with recent episodes of
star-formation. Moreover, an $age/\tau=4$ corresponds to a residual
SFR 2\% of the initial
SFR, for an exponential star formation history, as adopted in this paper.
We have checked that
around this typical value galaxies at $2\le z\le 3$ show a prominent
4000 \AA~and Balmer break, the typical signatures of an emerging old
stellar population.

The two parameters used in this context, the $age/\tau$ ratio and
the stellar mass, are subject to uncertainties and biases related to
the synthetic libraries used to carry out the fitting of the galaxy SEDs,
as we discuss here. In addition, the
stellar mass generally turns out to be the least sensitive to
variations in input model assumptions, and the extension of the
SEDs to mid-IR wavelengths with IRAC tends to reduce the formal uncertainties
on the derived stellar masses, as shown in \cite{massgoods}.

\begin{figure}
\includegraphics[width=9cm]{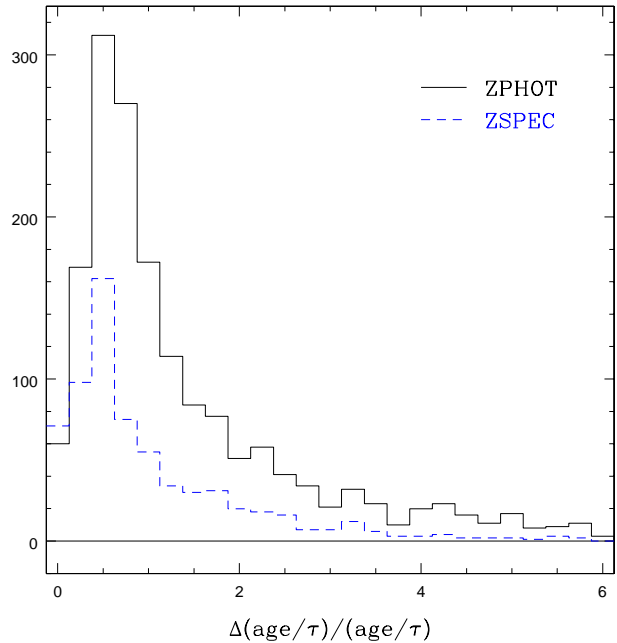}
\caption{
The relative uncertainties of the $age/\tau$ ratio. The dashed histogram
refers to galaxies with known spectroscopic redshift, whose $age/\tau$
parameter has a relative error of 40\%. The typical uncertainty
of this ratio for galaxies with a photometric redshift determination is
instead $\sim 50\%$.
}
\label{ttauunc}
\end{figure}

The uncertainties in the ratio $age/\tau$ 
are derived as follows: we compute the $1\sigma$ confidence level on
each estimate of $age$ and $\tau$ parameters by scanning the $\chi^2$
levels, allowing the redshift to change in case of objects with photometric
redshifts. The typical uncertainty, shown in Fig. \ref{ttauunc},
for the $age/\tau$ parameter is of the
order of 50\%, while the uncertainty for the
mass estimate within the GOODS-MUSIC sample is 40\%.
Though the age and $\tau$ parameters are individually poorly constrained and
highly degenerate, the ratio $age/\tau$ is comparatively better constrained.

Using the criterion based on the $age/\tau$ ratio on our Ks selected sample,
we identify 130 passively evolving galaxies ($age/\tau\ge 4$) in the
redshift range $1.4\le z\le 2.5$. 
Analogously, we find  508 young star forming galaxies (defined by the criterion
$age/\tau< 4$) in the same redshift range.

\begin{figure}
\includegraphics[width=9cm]{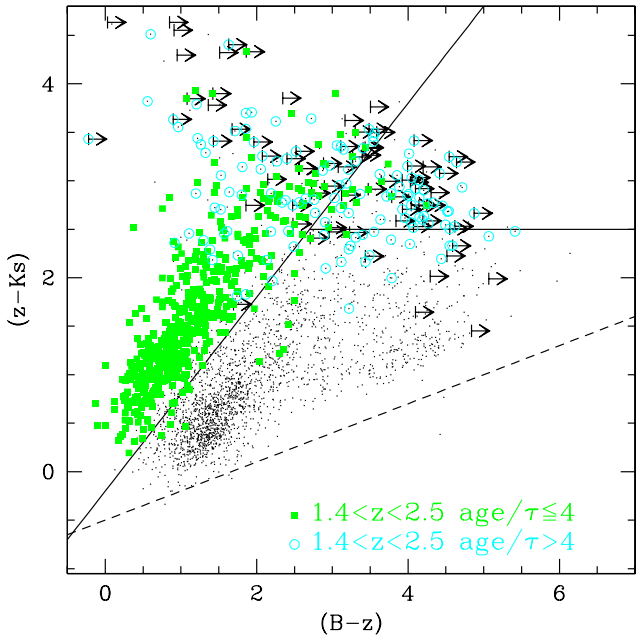}
\caption{
The $B-z$ vs $z-Ks$ colour for $Ks$--selected galaxies in the
GOODS-MUSIC sample, as in Fig. \ref{bzk}.
Green filled squares show the distribution of all $Ks$--selected galaxies with
redshift (spectroscopic or photometric) between 1.4 and 2.5, with young
stellar populations ($age/\tau\le 4$). Cyan circles are galaxies with
$1.4\le z\le 2.5$ and $age/\tau > 4$, a more efficient way to isolate
passively evolving galaxies than the original BzK-PE criterion.
}
\label{bzktau}
\end{figure}

Fig. \ref{bzktau} compares these galaxies selected through the
$age/\tau$ index with those identified by the BzK-PE and BzK-SF
criteria.  It is clear from this plot that the BzK and $age/\tau$ 
criteria are largely consistent.
In particular the BzK-SF criterion is very effective in recovering 
the starburst
phase of galaxies at $z\sim 2$ since it is able to select the large
majority (94\%) of galaxies with $age/\tau< 4$. On the other hand the
BzK-PE criterion is more incomplete (34\%), since it misses a great
fraction of passively evolving galaxies, having $age/\tau\ge 4$, especially at
faint Ks magnitude limits and for objects with extreme $z-Ks$ colours,
as shown in the previous section. In this case, the DRG criterion
turns out to have enhanced completeness in recovering passively evolving
galaxies, but also strong contamination
from dusty starbursts at $1<z<2$.
As a further validation of our approach based on the $age/\tau$ ratio,
we have selected all passively evolving galaxies with $age/\tau\ge 4$ and
$1.4\le z_{phot}\le 2.5$ in the BzK-SF locus and found one galaxy
(ID=9853 in GOODS-MUSIC, ID=139 in K20) with a spectroscopic redshift of
$z=1.553$ from the K20 survey (\cite{mignoli}) without [OII] line in emission
and with MgII line in absorption;
it is thus consistent with being an evolved galaxy.
Other galaxies
of this type are fainter, but the spectroscopic
identification is at the reach of deep or ultradeep spectroscopy carried out
by the GOODS (\cite{vanzella06}) and GMASS (\cite{gmass}) teams in the
CDFS field.

\begin{figure}
\includegraphics[width=9cm]{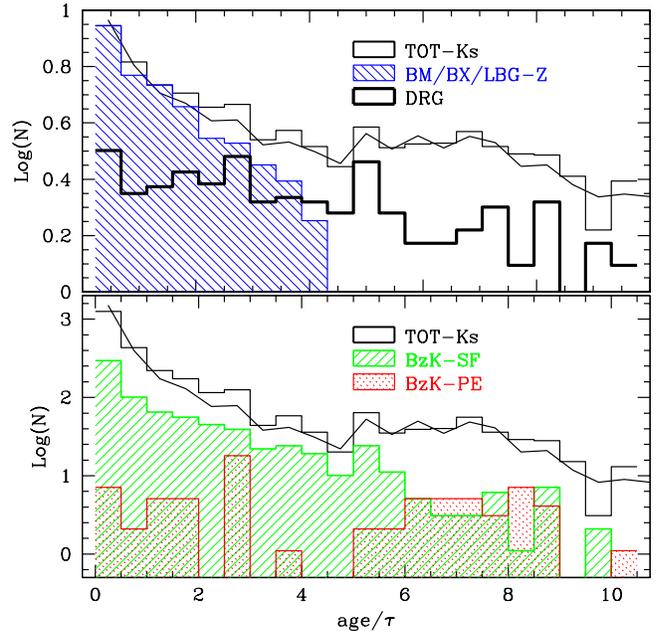}
\caption{Upper panel:
The distribution of the $age/\tau$ parameter for the total Ks--selected sample,
BM/BX/LBG-Z and  DRG. The continuous solid curve
shows the distribution of the $age/\tau$ ratio calculated using for
each galaxy the probability distribution function of this parameter. The curve
is consistent within the uncertainties with the histogram built using the
best fit values for each object. This shows that our approach, based on
the best fit quantities for $age/\tau$ is equivalent to the result that one
would obtain with a Monte Carlo approach.
Lower panel: The same of the upper panel for the total Ks--selected sample,
BzK-SF and BzK-PE galaxies.}
\label{histttau}
\end{figure}

We investigate also the distribution of $age/\tau$ for
different galaxy types.
The distribution of the $age/\tau$ ratio calculated using the probability
distribution function of this parameter for each galaxy,
is consistent within the uncertainties with the histogram built using the
best fit values for each object. This shows that our approach, based on
the best fit quantities for $age/\tau$ are equivalent to results that one
would obtain with a Monte Carlo approach.
Fig. \ref{histttau} confirms that BzK-SF and DRG galaxies are mixed
populations, and the distribution of the $age/\tau$ parameters is
an extended function, in which the relatively young objects dominate,
especially for the BzK-SF class. For DRGs the distribution is less steep,
with 110 objects with $age/\tau\le 4$ and 69 objects with $age/\tau >4$.
The BM/BX/LBGs are almost all at
$age/\tau\le 4$, confirming that they are predominantly young star forming
galaxies, while the BzK-PE class, despite the low statistic, shows
a bimodal distribution, with equal populations of young starbursts and
passively evolving objects. These facts strengthen our previous conclusions.

Note that from these figures (\ref{bzktau} and \ref{histttau}) we infer
a posteriori that the $age/\tau=4$ value is a good separator since an
apparent bimodality is evident in the $age/\tau$ distribution.
In particular, as shown in Fig. \ref{histttau},
the $age/\tau$ distribution for the Ks--selected galaxies flattens at
$age/\tau\ge 3.5-4.0$, indicating that a population of passively
evolving galaxies can be isolated with this proposed criterion.

\section{The stellar mass and mass density of high redshift galaxies}

The contribution of massive galaxies to the stellar mass function at
high redshifts has been already discussed by \cite{massgoods,caputi06}
and \cite{mclure}.  A
study of \cite{franceschini06}, based on detailed morphological
analysis with HST of a sample of galaxies detected at 3.6 $\mu$m in
the GOODS-South, indicates that the high mass tail of the stellar mass
function at $z\sim 1-2$ is due mainly to evolved galaxies. It is thus
interesting to derive the relative contribution of DRGs and BzK-PEs to
the stellar mass density (SMD) at high redshift. To provide a clear
comparison, we compute the SMD also for LBGs and BzK-SF galaxies,
deriving the contribution of star forming objects to the universal SMD
at high-z. Finally, we also derive the mass distribution and mass
density for $age/\tau\le 4$ and $age/\tau > 4$ galaxy samples, and
provide a global picture on the SMD for various galaxy types at $z\ge
1$.

The method that we applied to estimate the galaxy stellar masses
is described in Section 3.4.  A cautionary note on the stellar mass
derivation is necessary here, to remind that it always depends on the
assumed star--formation history. As discussed by many authors
(e.g. \cite{fontana04} and \cite{shapley}), the assumption of an
exponential decreasing SFR for all galaxies may provide biases in the
estimate of the stellar mass. In particular, it may lead to an
underestimate of the contribution of older stellar population in the
case of an actively star--forming galaxy, whose luminosity is
dominated by the glare of young stars.
The advantage of the exponential approach is that it
can be compared with many previous estimates.
In \cite{fontana04} we have shown
that it may lead to a maximal underestimate of the stellar mass by a
factor of 2, in the most extreme cases.
Another approach is the so-called maximally old model, in which the
near- and medium-IR SED is fitted with a burst of star formation at
$age\sim 0$, while the UV part of the spectrum is reproduced by a more
recent burst. This method recovers a larger stellar mass than that
evaluated by the best fit approach, but is based on extreme
assumptions, as shown in
\cite{fontana04}, and it is only useful to provide an upper limit
to the stellar mass assembled in a galaxy. We finally note that recent
papers (\cite{rettura}; \cite{erb}) show that the
dynamical mass derived from spectroscopy is consistent with the
stellar mass derived by the best fit approach, reinforcing the
robustness of our method.

\subsection{The observed mass distribution of high redshift galaxies}

\begin{figure*}
\includegraphics[width=9cm]{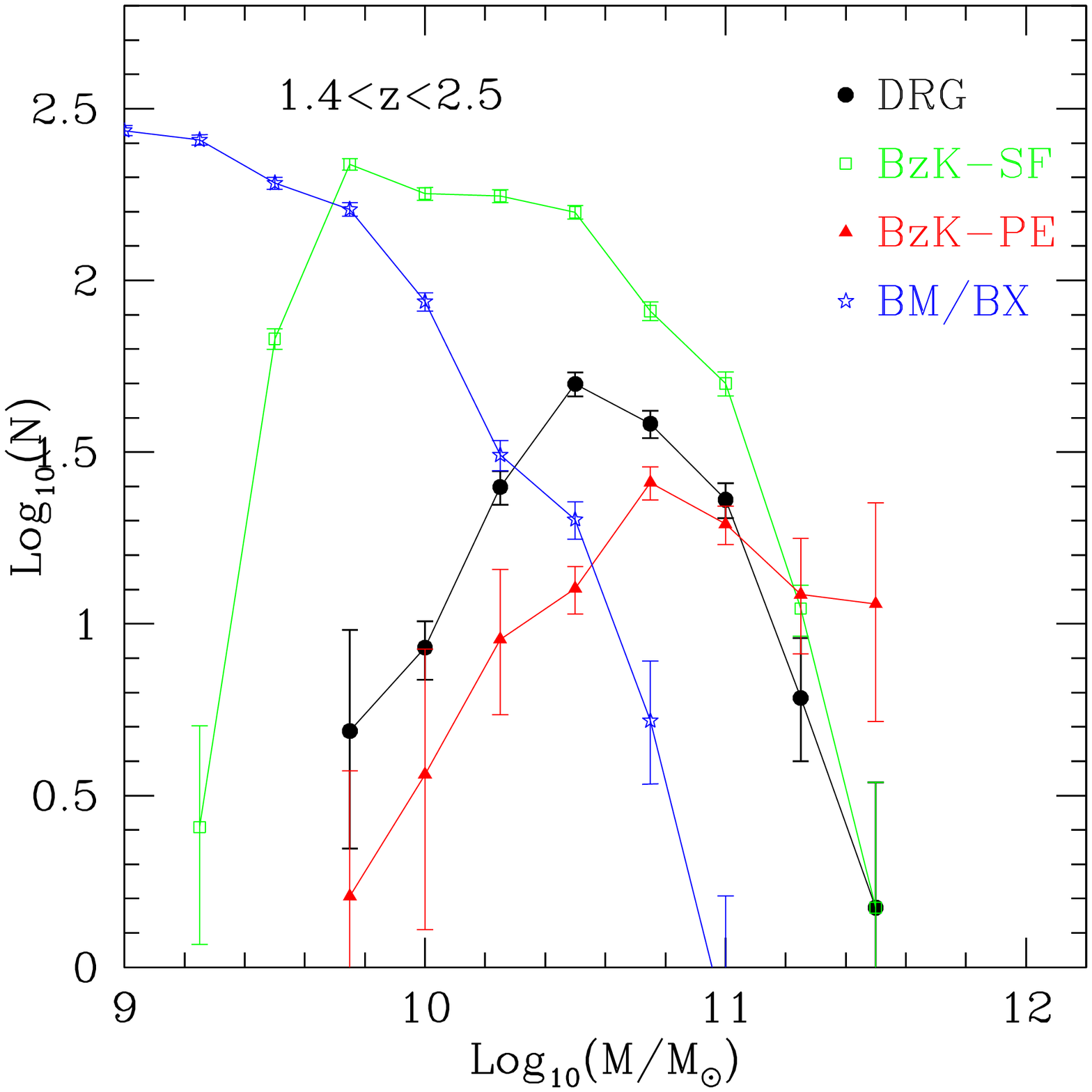} 
\includegraphics[width=9cm]{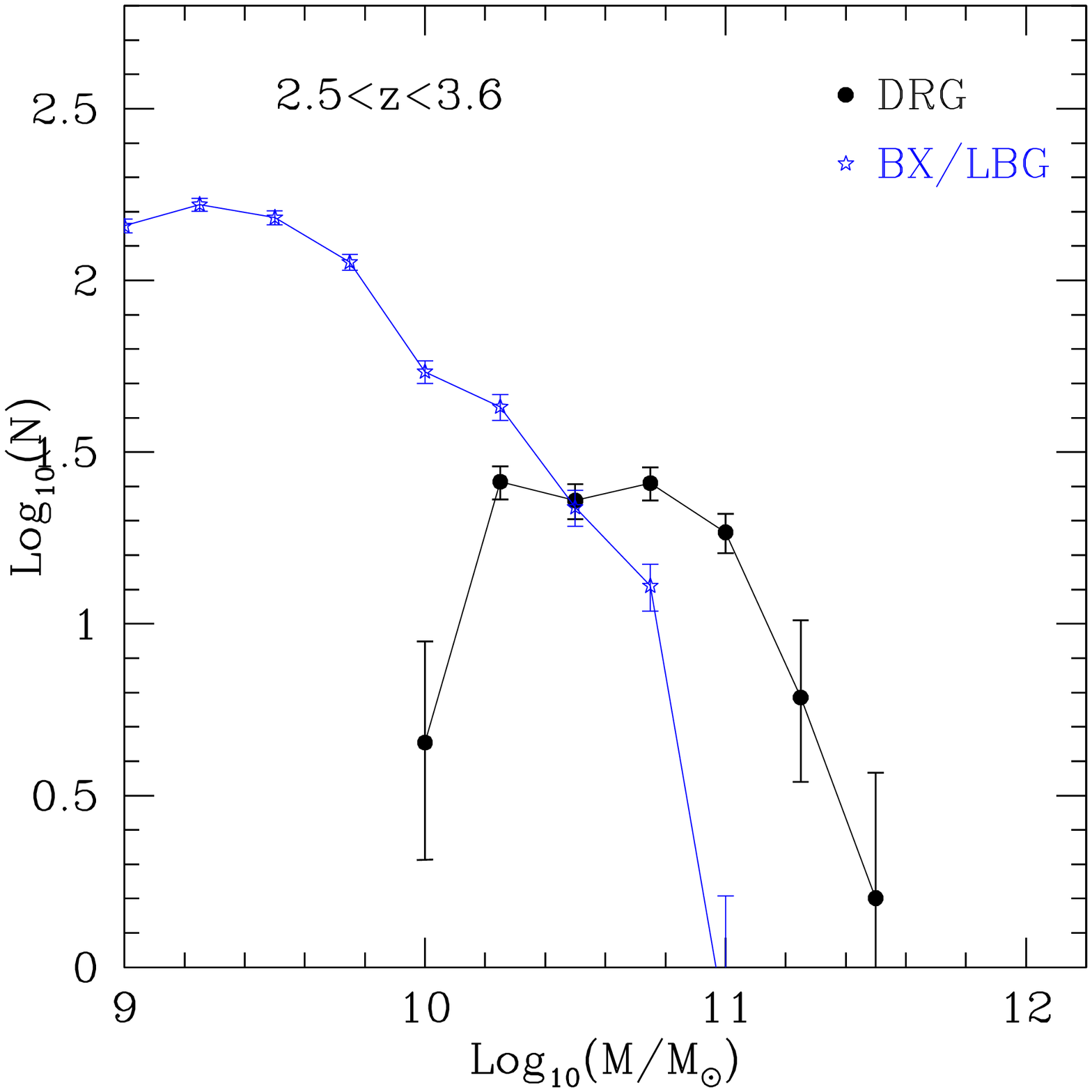}
\caption{
The distribution of stellar mass (in $M_{\odot}$ and logarithmic scale) for 
DRGs, BzKs (SF and PE) and BM/BX/LBGs in the redshift range $1.4\le z\le 2.5$
(left panel) and $2.5\le z\le 3.6$ (right panel).
Error bars represent uncertainties at 68\% confidence limit.}
\label{mhobs}
\end{figure*}

We first present the {\it observed} distribution of stellar
masses. Fig. \ref{mhobs} (left) represents the distribution of stellar
mass (note the logarithmic scale in the y--axis) for BM/BX/LBGs-Z,
BzKs and DRGs in the redshift interval $1.4<z<2.5$, while
Fig. \ref{mhobs} (right) refers to the redshift interval
$2.5<z<3.6$. Since the magnitude limit of the GOODS-MUSIC sample is
not unique both in z- and in Ks-band, the resulting mass distributions
are weighted for the different magnitude limits in the various
sub-areas (see Section 2 of this paper and also
\cite{grazian06a}).

It must be emphasised that the observed distributions do not
arise from a mass--limited sample, but rather from
magnitude--limited samples (in the $z$ or in the $Ks$--band),
and are therefore to be critically used. At high masses, all the samples
are complete (in stellar mass): all the conclusions
drawn on such samples at high masses are therefore robust. At low
masses, two different effects are in place. First, the $z$--selected
sample extends to fainter flux levels, and hence lower masses. In
addition, even within the $Ks$--selected samples (DRGs and BzKs),
star--forming spectral types extends to lower masses because their
typical $M/L$ ratio is lower than those of passively evolving
objects\footnote{We remind that the stellar mass of each galaxy is
derived through the mass-to-light ratio fitted by the spectral
synthesis model. Since evolved galaxies and star forming galaxies are
characterised by distinct $M/L$ ratios, a given cut in luminosity
translates in different cuts in mass for evolved and young stellar
populations}.  As described at length in \cite{massgoods}, a safe mass
threshold, i.e. the mass over which the samples are definitely ($\sim 100\%$)
complete in stellar mass, may be computed at each redshift using a maximally
old, single burst model. When applied to the present $Ks$--selected
sample, it turns out to be $\log(M_*/M_\odot) \simeq 10.6$ at $z\simeq
2$ and $\log(M_*/M_\odot) \simeq 10.85$ at $z\simeq 3$.
For the BM/BX/LBGs, selected from the $z$ band, we do not apply any
cut in stellar mass, since the relation between rest frame UV light
(redshifted in the optical bands) and the stellar mass is too
scattered at these redshifts for this type of galaxies.

The observed mass distributions are also sensitive to the redshift
selection function of different galaxy types. This selection
function is not known a priori, but from the observed redshift
distributions, which are not uniform over $1\le z\le 4$, as shown in
Fig. \ref{zhist}, it is clear that the selection functions of BzKs,
DRGs and BM/BX/LBGs are quite different.

We therefore emphasise that the results of this sub-section are
intended to provide a comparison of the contribution of
different samples of galaxies as provided by current methods and
surveys. A comparison of their overall impact on the global stellar mass
density is discussed instead in the following section.

With all these caveats in mind, we can compare the observed distributions
of stellar masses.  At $1.4<z<2.5$ (Fig. \ref{mhobs}, left), the fraction
of galaxies at the high-mass tail ($M\ge 10^{11} M_{\odot}$) is
dominated by DRGs and by BzK galaxies (both SF and PE), while LBGs
dominate the distribution at lower masses. Current surveys detect an
extended tail of low--mass star--forming galaxies (LBGs and BzK-SF)
that is missed if the mass budget at high-z is measured only using
DRGs. Such galaxies are much more numerous than the higher mass,
redder galaxies, and their contribution to the mass budget is not
negligible, as we shall discuss in the following.
At higher redshifts (Fig. \ref{mhobs}, right), the mass
distributions of both DRGs and BX/LBGs resemble those at lower redshift,
especially in the massive tail.

\begin{figure*}
\includegraphics[width=9cm]{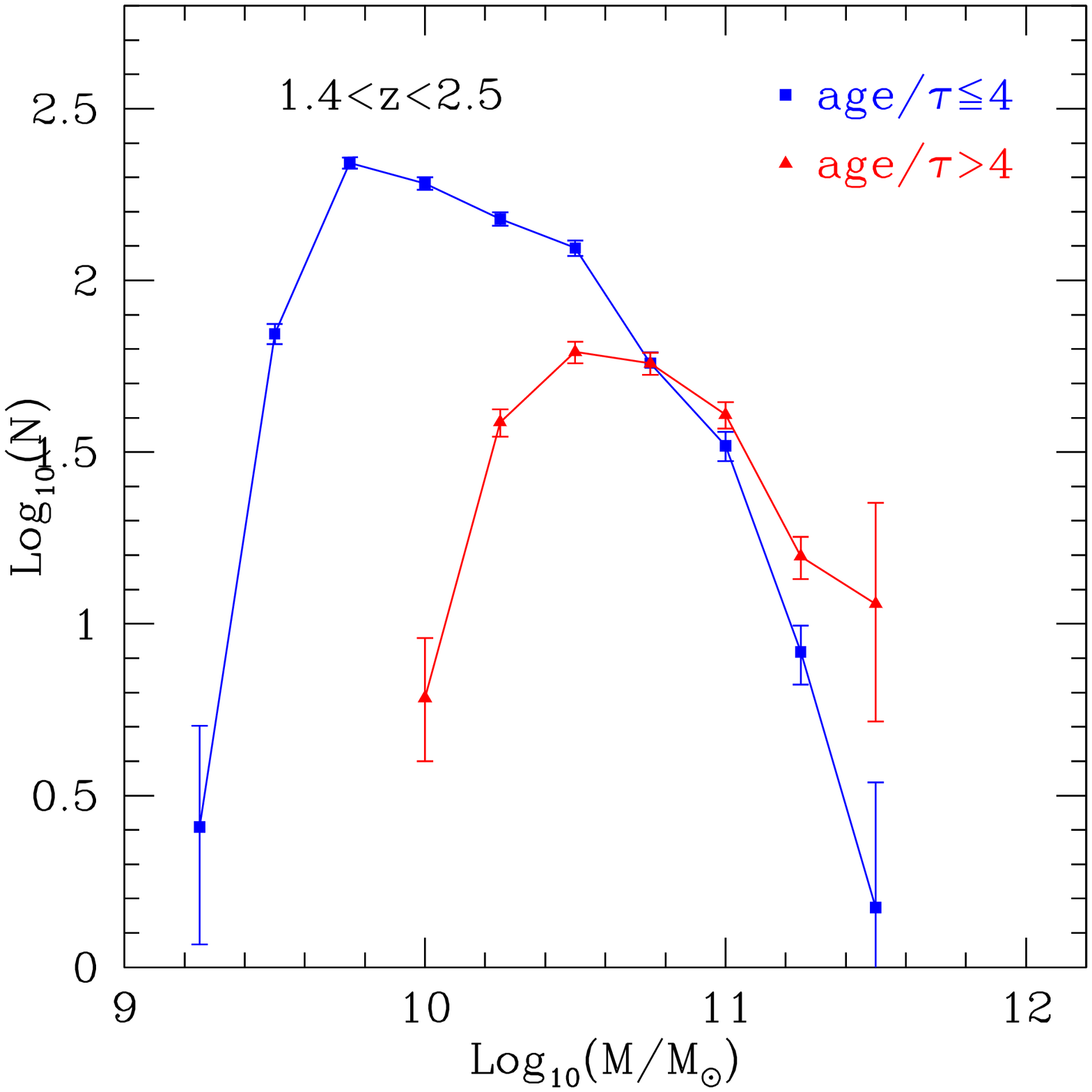}
\includegraphics[width=9cm]{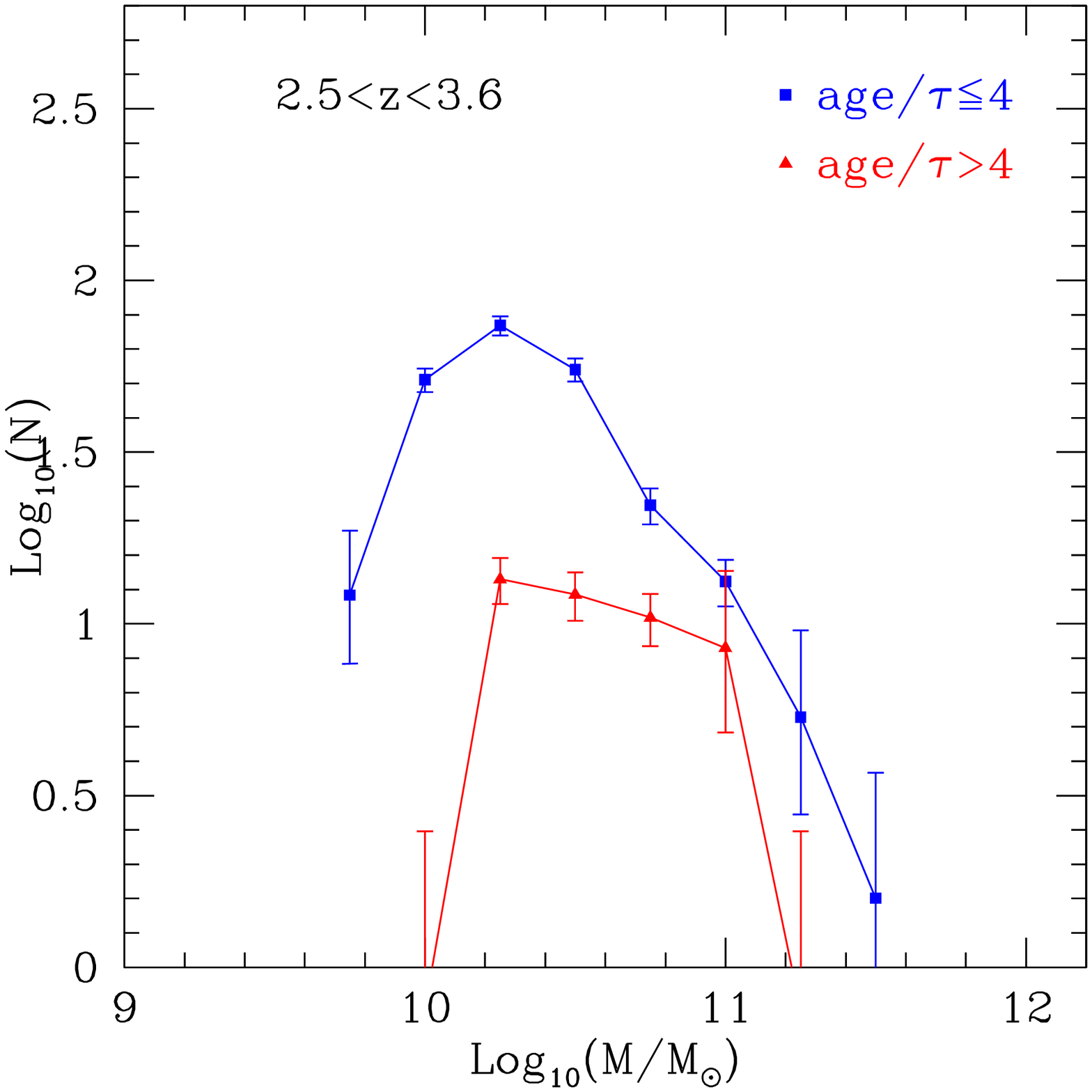}
\caption{
The distribution of stellar mass (in $M_{\odot}$ and logarithmic scale) for 
$age/\tau>4$ and $age/\tau\le4$ galaxies in the redshift range
$1.4\le z\le 2.5$ (left panel) and $2.5\le z\le 3.6$ (right panel).
Error bars represent uncertainties at 68\% confidence limit.}
\label{mhtau}
\end{figure*}

We also investigate the stellar mass distribution according to the
$age/\tau$ parameter: in Fig. \ref{mhtau} we plot the distribution of
stellar mass for $age/\tau >4$ and $age/\tau\le4$ galaxies in the
redshift range $1.4\le z\le 2.5$ (left panel) and $2.5\le z\le 3.6$
(right panel).

Fig. \ref{mhtau} shows that at $M\ge 10^{11} M_{\odot}$ and at
$z\sim 2$ the passively evolving galaxies are the dominant population,
while at $z\sim 3$ the high mass tail is mainly due to star forming
galaxies. We are thus witnessing the epoch of mass assembly (``upsizing'') of
passively evolving galaxies, confirming the strong density evolution of the
Stellar Mass Function for such galaxies at $z\ge 2$, as shown in
\cite{drory,franceschini06} and \cite{massgoods}.
On the other hand, the high mass tail of star forming galaxies
($age/\tau\le 4$) shows no evolution from $z\sim 2$ to $z\sim 3$, while
the low mass distribution experiences a strong negative evolution at
higher redshifts, yet an other indication of the so-called ``downsizing''.

We finally note that the mass distribution of passively evolving
galaxies extends somewhat below the mass cut of the BzK-PE.  The fall
down of the mass distribution for BzK-PE is indeed due to the
selection effect already described in Sect. 3.5, by which the BzK
criterion turns out to be incomplete in the selection of passively
evolving galaxies at faint $Ks$ magnitudes.

\subsection{The stellar mass density of high redshift galaxies}

The Stellar Mass Density analysis for high-z galaxies is a more
appropriate description
than the mass distribution presented in the previous paragraph, since
it takes into full account volume effects.
In addition, and even more important, it allows to
compare the contribution of the different classes of galaxies, as
selected with the various methods described here, with the total
stellar mass density, as obtained by integrating the stellar mass
function at the corresponding redshifts.

The contribution to the total observed SMD at $2<z<3$ and
$M>10^{11}M_{\odot}$ has been derived by \cite{vandokkum06} and is
77\% and 17\% for DRGs and LBGs, respectively. In a similar way,
\cite{rudnick} measured the SMD of DRGs at $z\sim 2.8$ over different lines
of sight and found that these galaxies contribute to 64\% of the observed
SMD at this redshift.

We compute the contribution to the SMD for LBGs and DRGs in the same
mass and redshift range of \cite{vandokkum06} and we find that
BM/BX/LBGs-Z recover 9\% of
the total observed SMD, while DRGs make up 64\% of the observed
SMD. This result agrees quite well with that of \cite{vandokkum06} and
\cite{rudnick}.

We can expand this analysis using our deeper GOODS-MUSIC catalog: we
derive the mass density $\rho^\ast_M$ for DRGs, BzKs and BM/BX/LBGs
separately as a function of redshift, and compare it with the observed
$\rho^\ast_M$ for the Ks selected sample and with the total value
obtained integrating the fitted mass function over the mass range
$10^8-10^{12}M_\odot$ (\cite{massgoods}). We then repeat the same
analysis on the $age/\tau$ selected samples. 

We have divided the DRGs and BM/BX/LBGs samples in three redshift
intervals, following the BM, BX and LBG  redshift
distributions. The BzK contribution to the stellar mass density is
computed only in the redshift interval $1.4\le z\le 2.5$ where the BzK
criterion is effective. The $age/\tau$ selected samples are
arbitrarily divided into two redshift intervals, $1.4\le z\le 2.5$ and
$2.5\le z\le 3.6$; the first redshift bin is chosen so that direct
comparison with the BzK criterion is possible, while the second bin is
almost identical to the high redshift DRG and LBG redshift
distributions as in Fig. \ref{mhobs} and \ref{mhtau}.
Fig. \ref{massdens} shows the SMD for DRGs, LBGs and BzK galaxies at
different redshifts, while Tab. \ref{mdenstab} contains the SMDs for
these galaxy types. The error bars of the SMD have been computed
with a full Monte Carlo simulation where we take into account the
redshift probability distribution for each galaxy in the sample.

These numbers are to be compared with the total mass density obtained
by integrating between $10^8$ and $10^{12}M_{\odot}$ the global Galaxy
Stellar Mass Function (GSMF) observed in the same GOODS--MUSIC sample
(\cite{massgoods}). Clearly, these fractions are relatively small since
the lower integration limit for the total SMD is far below the
observational limits for both the Ks and the z--selected samples.
Finally we repeat the same calculations for $M>10^{11}M_{\odot}$.

The value of the total (integrated) stellar mass density depends on
the slope of the GSMF at the low mass side, which is presently not
known at these redshifts. The Schechter slope index $\alpha$ is
observed to remain considerably flat, changing from $\alpha=-1.18$ at
$z=0$ to $\alpha\simeq -1.3$ at $z\simeq 1.3$ (\cite{massgoods}). As a
first estimate, we shall assume that this trend of slowly decreasing
$\alpha$ will continue up to $z\simeq 3$ (where $\alpha$ would be
$-1.54$), as implicit in the Schechter parametric representation of
the GSMF provided in \cite{massgoods}. We shall discuss in the
following the impact of releasing this assumption on the slope of the
GSMF.

In the first redshift bin, $1.4\le z\le 2.2$, DRGs recover 21\% of the
total mass density, similar to the contribution of BM-Z galaxies, that
is 14\% (see Tab. \ref{mdenstab}). BzK-SF proves to be a more
efficient criterion recovering 75\% of the total SMD, while BzK-PE
provide around 19\% of the total mass density.  Consistently with our
previous discussion of the relative efficiency of the selection
criteria, the BzK criteria turn out to be more efficient than BM-Z or
DRGs for deriving the total SMD.  Moreover, BzK-SF galaxies contribute
to 75\% of the total SMD, indicating that at these redshift a
considerable fraction of the mass budget can be found in star forming
galaxies, possibly obscured by dust.

In the higher redshift bins, BX/LBG-Z galaxies contribute to 58\% and
48\% of the total SMD at $2.2\le z\le 2.8$ and $2.8\le z\le 3.7$,
respectively, while DRGs recover 37 and 28\% in the same redshift
intervals.  This implies that BX/LBG-Z galaxies cannot be neglected
when the total SMD is investigated. Of course, limiting the analysis
to the high mass tail of the galaxy mass function, i.e. masses larger
than $10^{11}M_{\odot}$, the contribution of DRGs is predominant, as
we can see in the lower part of Tab.\ref{mdenstab} and in agreement
with \cite{vandokkum06} and \cite{rudnick}.

To further check these results we have derived the contribution of
passively evolving and star forming galaxies selected according to the
age/$\tau$ parameter: star forming galaxies provide a considerable
fraction of the total mass budget, in particular at redshift $\sim$3.
The contribution of passively evolving galaxies ($age/\tau>4$) at
$1.4<z<2.5$ is 32\%. At higher redshift the contribution of
such galaxies becomes even lower, only 13\%.  

These results are in practice obtained by integrating the observed
mass distributions of Fig. \ref{mhobs} and Fig. \ref{mhtau} (with the
proper volume element) and comparing them with the integrated GSMF. As
mentioned before, the observed distributions of Fig. \ref{mhobs} and
Fig. \ref{mhtau} are biased by selection effects against low mass, red
galaxies, while the total SMD depends on the assumed slope of the
GSMF. Because of these systematics, it is possible that
the observed distributions of Fig. \ref{mhobs} and
Fig. \ref{mhtau} are only the results of a selection effect, and that red
(or passively evolving galaxies) keep dominating the stellar mass
budget even at lower masses, overpredicting the total SMD. This would
imply that the slope of the
GSMF becomes significantly steeper than what assumed here at high
redshift.
In the following subsection, we try using the same GOODS--MUSIC data
as a first attempt in this direction.

\subsection{Looking for faint DRGs in the Ks-band}

As discussed above, the DRGs and in general all red
galaxies are limited to high masses simply due to their typically
higher $M/L$: pushing our analysis to one magnitude deeper than the
actual magnitude limits, it is possible to investigate the
existence of a population of faint/low-mass DRGs, whose contribution
to the SMD could overcome that of LBGs.

\begin{figure}
\includegraphics[width=9cm]{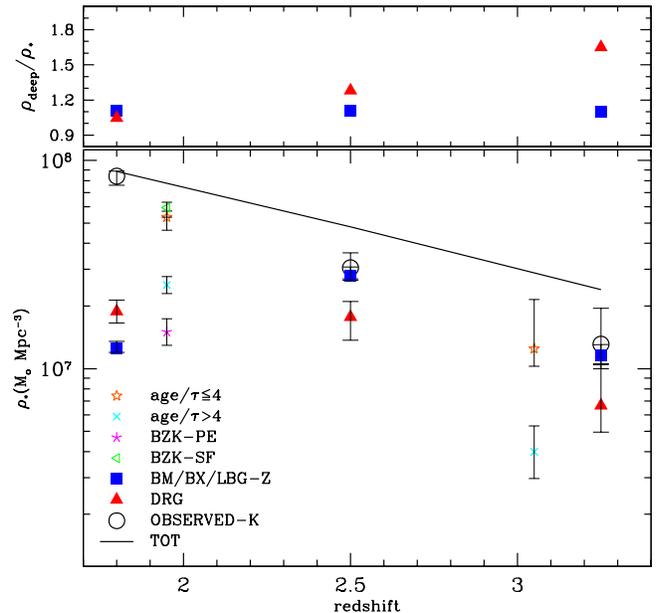}
\caption{
{\em Lower panel:}
the stellar mass density of the observed Ks--selected sample (circles), DRGs 
(filled triangles) and LBGs (filled squares) for the redshift bins where
the BM,  BX and LBG selections are efficient. DRGs sample a considerable 
fraction of the stellar mass density of the Universe at $z\sim 2.5$, 
but LBGs are complementary to DRGs in recovering the total 
$\rho^\ast_M$. The BzK-SF criterion (void triangles) is very 
efficient in the lower redshift bin $1.4\le z\le 2.5$, surpassing 
also the contribution of the DRG sample. The contribution of BzK-PE
galaxies (asterisk) is slightly reduced, due to incompleteness 
of this selection criterion. Contrarily, the selection $age/\tau>4$
(crosses) recovers a higher stellar mass density, which is the contribution of
true passively evolving galaxies at $z\sim 2$. The dominant sample at $z\ge 1$
is however that selected by $age/\tau\le 4$ (stars). The continuous line
shows the mass density obtained integrating the total galaxy stellar 
mass function between $10^8$ and $10^{12} M_\odot$.
{\em Upper panel:}
the magnitude limits both in the
$z$ and in the $Ks$ bands are then lowered down of one magnitude in order to
try to derive the contribution of faint galaxies to the mass density at
high-z ($\rho_{deep}$). At $z\sim 3$ the increase on SMD for faint DRGs
is larger by a factor $\sim 1.6$ than the increase due to faint LBGs.
}
\label{massdens} 
\end{figure}

To push the present sample toward lower mass galaxies, we derive the
stellar mass density shifting by one magnitude both the actual
magnitude limits of the survey, in the z as well as in the Ks bands,
(corresponding to 90\% completeness all over the GOODS-South field, as
described in Section 2).  In this case, the distribution is only a
lower limit to the number density of galaxies at low masses, while it
is well constrained at the high mass tail. If we compute the SMD for
LBGs and DRGs using a mass limit $\sim$2.5 times lower (corresponding
to one magnitude deeper in the z and Ks bands, or alternatively to 75\%
completeness) than the
conservative one, we derive a SMD higher both for LBGs and DRGs, as we
expect. The differential increase however is not constant, as shown in
Fig. \ref{massdens} (upper panel), but DRGs gain more ($\sim 1.6$ times)
than LBGs in terms of SMD, especially at $z\sim 3$. This provides an
indication for the existence of a numerous population of faint and
less massive DRGs than those observed today, although it is still unclear
whether they could overcome in number and mass density the
contribution of LBGs at high redshift. 

The mass distribution of DRGs thus seems to be continuous and not
limited to the high mass tail. A survey much deeper than GOODS, like
the ultra-deep FIRES Ks-band observation of the HUDF (\cite{udf}, Labb\'e
et al., in prep.),
will probably be able to constrain the number density of low-mass DRGs
in this redshift range.

\begin{table}
\caption[]{Stellar mass density for DRGs, BM/BX/LBGs and BzK galaxies}
\begin{tabular}{lccc}
\hline
\hline
Sample & $z\sim 1.8$ & $z\sim 2.5$ & $z\sim 3.3$ \\
\hline
$Log~\rho^\ast_M$ & & & \\
\hline
TOT            &   7.947  &   7.681    &   7.380 \\
OBSERVED-Ks    &   7.924  &   7.485    &   7.118 \\
DRG            &   7.275  &   7.248    &   6.823 \\
BM/BX/LBG-Z    &   7.096  &   7.440    &   7.064 \\
BzK-SF         &   7.774  &            &    --   \\
BzK-PE         &   7.177  & --         &    --   \\
$age/\tau>4$   &   7.402  & --         &   6.601 \\
$age/\tau\le4$ &   7.728  & --         &   7.096 \\
\hline
Type/TOT & & & \\
\hline
DRG            &     0.21   &   0.37  &     0.28  \\
BM/BX/LBG-Z    &     0.14   &   0.58  &      0.48 \\
BzK-SF         &     0.75   &     --   &        --    \\
BzK-PE         &     0.19   &     --   &        --    \\
$age/\tau>4$   &     0.32   &     --   &      0.13  \\
$age/\tau\le4$ &     0.68   &     --   &      0.42  \\
\hline
Type/TOT($M>10^{11}M_{\odot}$) & & &   \\
\hline
DRG            &    0.30     &     0.77  &   0.79   \\
BM/BX/LBG-Z    &    0.01     &     0.09   &  0.10   \\
BzK-SF         &    0.57     &      --      &      --     \\
BzK-PE         &    0.37     &      --      &      --     \\
$age/\tau>4$   &    0.55     &      --      &   0.25   \\
$age/\tau\le4$ &    0.45     &      --      &   0.75   \\
\hline
\hline
\end{tabular}
\label{mdenstab}
\\
$\rho^\ast_M$ is the stellar mass density in $M_\odot Mpc^{-3}$. The
three redshift intervals, $1.4\le z\le 2.2$, $2.2\le z\le 2.8$ and
$2.8\le z\le 3.7$, correspond to the
BM, BX and LBG selection criteria, respectively.
$\rho^\ast_M({TOT})$ is the mass density obtained 
integrating the stellar mass function between $10^8$ and $10^{12}
M_\odot$. The
BzK criterion is effective only in the redshift range $1.4\le z\le 
2.5$, and for the higher redshift bins it is not possible to compute 
its stellar mass density. The SMD for galaxies divided according to their
$age/\tau$ parameter is computed in the redshift intervals $1.4\le z\le 2.5$
and $2.5\le z\le 3.6$. The total SMD in these redshift ranges are
$10^{7.898}$ and $10^{7.471}$, and these two quantities are used to compute
the relative contribution of BzK galaxies and of objects selected by their
$age/\tau$ ratio.
\end{table}


\section{Discussion and Conclusions}

In this work  we have used the GOODS-MUSIC sample to compare
the colour criteria commonly used for selecting high redshift galaxies,
and to estimate their contribution to the universal stellar mass budget.

The GOODS--MUSIC sample is particularly suited for this exercise,
since it provides a reasonable statistics (nearly 3000 galaxies in the
$Ks$--selected sample and more than 9000 in the $z$--selected one), a
14 bands wavelength coverage that allows a direct application of most
colour selection criteria and well tested, accurate photometric redshifts. 
Most important,
it includes the $3.6-8\mu m$ Spitzer observations of the complete data
set, that are necessary to provide reliable estimates of the stellar
mass for galaxies at relatively high redshift (\cite{massgoods}).

We have initially selected samples of BM/BX/LBGs, DRGs and BzK
galaxies, discussing the overlap between the various sets and the
limitations of each criterion.  As expected from previous works, the
selection of galaxies according to the BM/BX/LBG criteria is sensitive
to moderately obscured star-forming galaxies, missing the dusty starburst
objects. DRGs, instead, are less sensitive to dust obscuration
effects, but comprise a mix of two populations, the old/evolved
galaxies and the dusty starbursts at intermediate/high redshifts. The
BzK criterion is highly efficient in the redshift range $1.4\le z\le
2.5$, but when galaxies start to become faint in the Ks band and red
in the $z-Ks$ colour, it is difficult to distinguish between
star-forming and evolved galaxies, resulting in an underestimation of
the passively evolving population.

To better separate actively star--forming galaxies from passively
evolving ones, we have then applied a physical criterion, based
on the ratio between the age of the stellar component and the
star--forming timescale, both derived from the best--fitting models
applied to the full SED of each galaxy.  We show that the ratio
age$/\tau$ is a relatively well constrained parameter, and we adopt a
threshold age$/\tau>4$ to separate passively evolving galaxies from
star forming ones.

We have then analysed the observed distribution of stellar masses associated
to each selection criterion. At $1.4<z<2.5$, the fraction of galaxies
in the high-mass tail ($M\ge 10^{11} M_{\odot}$) is dominated by DRGs
and by BzK galaxies (both SF and PE), while BM/BX and BzK-SF dominate
the distribution at lower masses. The same kind of bimodality is
apparently in place at $2.5<z<3.6$, where DRGs make the high mass tail
of the distribution and BX/LBGs the low mass side. Clearly, the lack of
red galaxies with low stellar mass
results from the selection criteria adopted: since the
$z$--selected sample extends to fainter fluxes than the $Ks$--selected
one, and given that the $M/L$ ratio of red galaxies is larger than
that of blue galaxies, current samples are biased against the
detection of low mass red galaxies.

If we look at the intrinsic properties, it is remarkable that at
$z\simeq 2$ passively evolving galaxies, selected with $age/\tau>4$,
exist in a large number, such that they are slightly dominant at the
highest masses ($M\ge 10^{11} M_{\odot}$) over the star--forming
population. However, they show a significant negative evolution with
redshift, with a more modest contribution to the high mass tail at
$z\simeq 3$. This transition suggests that the epoch from $z\simeq 2$
to 3 is crucial in the assembly of passively evolving
galaxies.

We have then derived the contribution of each selected sample to the
Stellar Mass Density (SMD) at various redshifts. This is compared with
the total SMD, as obtained by
integrating the Galaxy Stellar Mass Function of \cite{massgoods} from
$10^8 M_\odot$ to $10^{12} M_\odot$. Such a comparison is based on the
assumption that the slope of the Galaxy Stellar Mass Function at high
redshift remains relatively flat, as inferred from an the extrapolation of
its evolution at lower redshifts (\cite{massgoods}).  Under this
assumption, the overall contribution of BM/BX/LBGs to the total SMD
from $z\sim 1.5$ to $z\sim 3.5$ is comparable to or even higher than
that of DRGs, since they outnumber the DRGs at low masses. Clearly, if
only the high mass tail is considered, DRGs are the dominant
population.  The importance of star forming galaxies is even more
noticeable if one considers the BzK-SF sample, which is dominating the
SMD at $1.4\le z\le 2.5$, since it recovers both moderately obscured
and dusty young galaxies.

If this picture is correct, it presents a similarity and a difference
with the present-day Universe. The similarity is that a bimodality
seems to exist, with star--forming galaxies dominating the low mass
population and passively evolving galaxies dominating the high mass
tail. At variance with the local Universe, however, the integrated
contribution of passively--evolving galaxies is lower than that of
star--forming ones. These two clues provide a different
view of the so called ``downsizing'' scenario.

However, there is the possibility that a large population of
low--mass, intrinsically red galaxies exists, still
undetected in current K--selected surveys. Such a population could in
principle produce a Galaxy Stellar Mass Function much steeper than
assumed here, and would significantly change the relative contribution
of different galaxy types to the SMD. Detecting the existence of such
a population, and constraining its nature (passively evolving
vs. star--forming) would be very important to understand the physical
processes in high redshift galaxies, since in theoretical scenarios
feedback effects and
star--formation histories are strongly dependent on the halo mass.
Only much deeper IR--selected surveys
will be able to discover whether the contribution of DRG and other red
galaxies to the SMD at $z\ge 2$ is actually higher than what observed
with present data: DRGs indeed could eventually overcome the BM/BX/LBG
SMD when their contribution is integrated till the very low mass tail
of the mass distribution.

\begin{acknowledgements}
It's a pleasure to thank the GOODS Team for providing all the imaging material
available worldwide. Observations have been carried out using the Very Large
Telescope at the ESO Paranal Observatory under Program IDs LP168.A-0485 and
ID 170.A-0788 and the ESO Science Archive under Program IDs 64.O-0643,
66.A-0572, 68.A-0544, 164.O-0561, 163.N-0210 and 60.A-9120.
We are grateful to the referee for useful, detailed and constructive comments.
AG warmly thanks E. Daddi for useful discussions about the nature of faint
red BzK galaxies. 
\end{acknowledgements}

\end{document}